\documentclass[aps,reprint,twocolumn,amsmath,amssymb,superscriptaddress,prb]{revtex4-1}

\usepackage{graphicx}
\usepackage{dcolumn}
\usepackage{bm}
\usepackage{units}
\usepackage{upgreek}
\usepackage{microtype}
\usepackage{hyperref}
\usepackage{multirow}

\begin{document}
\title{Optical Phonons in Methylammonium Lead Halide Perovskites and Implications for Charge Transport}

\author{Michael Sendner}
\affiliation{InnovationLab, Speyerer Str.\ 4, 69115 Heidelberg, Germany}
\affiliation{Kirchhoff Institute for Physics, Heidelberg University, Im Neuenheimer Feld 227, 69120 Heidelberg, Germany}
\author{Pabitra K.\ Nayak}
\affiliation{Clarendon Laboratory, Department of Physics, University of Oxford, Oxford OX1 3PU, UK}
\author{David A.\ Egger}
\affiliation{Department of Materials and Interfaces, Weizmann Institute of Science, Rehovoth 76100, Israel}
\author{Sebastian Beck}
\affiliation{InnovationLab, Speyerer Str.\ 4, 69115 Heidelberg, Germany}
\affiliation{Kirchhoff Institute for Physics, Heidelberg University, Im Neuenheimer Feld 227, 69120 Heidelberg, Germany}
\author{Christian M\"{u}ller}
\affiliation{InnovationLab, Speyerer Str.\ 4, 69115 Heidelberg, Germany}
\affiliation{Kirchhoff Institute for Physics, Heidelberg University, Im Neuenheimer Feld 227, 69120 Heidelberg, Germany}
\affiliation{Institute for High Frequency Technology, TU Braunschweig, Schleinizstr.\ 22, 38106 Braunschweig, Germany}
\author{Bernd Epding}
\affiliation{InnovationLab, Speyerer Str.\ 4, 69115 Heidelberg, Germany}
\affiliation{Kirchhoff Institute for Physics, Heidelberg University, Im Neuenheimer Feld 227, 69120 Heidelberg, Germany}
\affiliation{Institute for High Frequency Technology, TU Braunschweig, Schleinizstr.\ 22, 38106 Braunschweig, Germany}
\author{Wolfgang Kowalsky}
\affiliation{InnovationLab, Speyerer Str.\ 4, 69115 Heidelberg, Germany}
\affiliation{Kirchhoff Institute for Physics, Heidelberg University, Im Neuenheimer Feld 227, 69120 Heidelberg, Germany}
\affiliation{Institute for High Frequency Technology, TU Braunschweig, Schleinizstr.\ 22, 38106 Braunschweig, Germany}
\author{Leeor Kronik}
\affiliation{Department of Materials and Interfaces, Weizmann Institute of Science, Rehovoth 76100, Israel}
\author{Henry J.\ Snaith}
\affiliation{Clarendon Laboratory, Department of Physics, University of Oxford, Oxford OX1 3PU, UK}
\author{Annemarie Pucci}
\affiliation{InnovationLab, Speyerer Str.\ 4, 69115 Heidelberg, Germany}
\affiliation{Kirchhoff Institute for Physics, Heidelberg University, Im Neuenheimer Feld 227, 69120 Heidelberg, Germany}
\affiliation{Centre for Advanced Materials, Heidelberg University, 69120 Heidelberg, Germany}
\author{Robert Lovrin\v{c}i\'{c}}
\email[Author to whom correspondence should be addressed. Electronic mail: ]{r.lovrincic@tu-braunschweig.de}
\affiliation{InnovationLab, Speyerer Str.\ 4, 69115 Heidelberg, Germany}
\affiliation{Institute for High Frequency Technology, TU Braunschweig, Schleinizstr.\ 22, 38106 Braunschweig, Germany}

\date{\today}

\begin{abstract}
Lead-halide perovskites are promising materials for opto-electronic applications. Recent reports indicated that their mechanical and electronic properties are strongly affected by the lattice vibrations. Herein we report far-infrared spectroscopy measurements of CH$_{3}$NH$_{3}$Pb(I/Br/Cl)$_{3}$ thin films and single crystals at room temperature and a detailed quantitative analysis of the spectra. We find strong broadening and anharmonicity of the lattice vibrations for all three halide perovskites, which indicates dynamic disorder of the lead-halide cage at room temperature. We determine the frequencies of the transversal and longitudinal optical phonons, and use them to calculate the static dielectric constants, polaron masses, electron-phonon coupling constants, and upper limits for the phonon-scattering limited charge carrier mobilities. Our findings place an upper limit in the range of $\unit[200]{cm^{2}V^{-1}s^{-1}}$ for the room temperature charge carrier mobility in MAPbI$_{3}$ single crystals, and are important for the basic understanding of charge transport processes and mechanical properties in metal halide perovskites.
\end{abstract}

\maketitle

\section{Introduction}
Lead-halide perovskites, with the general chemical structure {\it ABX}$_{3}$, have recently reemerged as promising materials for opto-electronic applications.\cite{zhang_metal_2016,kojima_organometal_2009,lee_efficient_2012} The most prominent candidates to date for perovskite-based solar cells are the hybrid organic-inorganic perovskites {\it ABX}$_{3}$\citep{mitzi_synthesis_1999}, where {\it A} is the organic cation, $B=\mathrm{Pb}$, and {\it X} is the halide. The most widely studied examples in this material class are the methylammonium (MA) lead halides\cite{weber_d_ch_1978} (MAPb(I/Br/Cl)$_{3}$). Their bandgaps range from approximately $\unit[1.6]{eV}$ (MAPbI$_{3}$) to $\unit[3.0]{eV}$ (MAPbCl$_{3}$).\cite{liu_two-inch-sized_2015} They exhibit structural phase transitions in an orthorhombic-tetragonal-cubic sequence, with MAPbI$_{3}$ being tetragonal at room temperature, and MAPbBr$_{3}$ and MAPbCl$_{3}$ being both cubic.\cite{stoumpos_semiconducting_2013} As these phase transitions can occur close to room temperature, structural fluctuations of the lead-halide cage have potentially strong impact on the opto-electronic properties.\cite{yaffe_nature_2016}

Long charge carrier diffusion lengths exceeding $\unit[1]{\upmu m}$ have been reported for MAPbI$_{3}$ and MAPbBr$_{3}$ single crystals and thin films.\cite{shi_low_2015-1,stranks_electron-hole_2013-1,Chen2016} The Urbach energies in these materials are $\sim\unit[15]{meV}$, comparable to the values for typical inorganic crystalline semiconductor materials like GaAs,\cite{de_wolf_organometallic_2014} suggesting a low prevalence of bandgap states, which often lower the carrier mobilities in semiconductors. Despite the low number of band gap states and the often observed high structural order of the cornor sharing lead halide octahedras, the charge carrier mobilities are modest, especially in view of the low effective masses.\cite{brenner_are_2015,brenner_hybrid_2016} Moreover, the thermal conductivity of MAPbI$_{3}$ has recently been calculated to be ultralow.\cite{wang_anisotropic_2016,Pisoni2014} All of these properties are intimately connected to the nature of the lattice vibrations in these materials. In particular, the modest charge carrier mobilities have been explained by either predominant electron scattering from acoustic\cite{milot_temperature-dependent_2015,oga_improved_2014,savenije_thermally_2014,yi_intrinsic_2016} or optical phonons\cite{la-o-vorakiat_phonon_2016,menendez-proupin_nonhydrogenic_2015-1,wright_electron-phonon_2016} or the formation of polarons\cite{zhu_charge_2015,yi_intrinsic_2016,neukirch_polaron_2016-1,nie_light-activated_2016,Soufiani2015}. Furthermore, recently the influence of spin-orbit coupling on the temperature dependence of the mobility was shown theoretically.\cite{Yu2016} To better understand such fundamental material properties, detailed knowledge of the phonon spectra is mandatory. Such can in principle be obtained from Raman measurements, which for MAPbX$_{3}$ are, however, experimentally challenging due to the ease of beam damage under laser irradiation. Far infrared (IR) spectroscopy circumvents the beam damage problem and reveals information complementary to Raman spectroscopy. Several far-IR studies of MAPbX$_{3}$ have appeared over the last two years.\cite{brivio_lattice_2015-1,la-o-vorakiat_phonon_2016,perez-osorio_vibrational_2015} P\'{e}rez-Osorio et al. measured the frequency of the transversal optical (TO) phonons and theoretically predicted a strong LO-TO splitting for the strongest vibration.\cite{perez-osorio_vibrational_2015} Wright et al. estimated the longitudinal optical (LO) phonon frequencies indirectly via temperature dependent photoluminescence measurements ($\omega_{\mathrm{LO}}$=11.5 and 15.3\,meV, 92 and 122\,cm$^{-1}$).\cite{wright_electron-phonon_2016} However, a direct experimental determination of the LO phonon frequency is currently missing. In view of the fact that LO phonons can interact with charge carriers, it is important to understand charge transport close to room temperature.

Herein we report far-IR spectra of the three methylammonium lead halide perovskite (MAPb(I/Br/Cl)$_{3}$) thin films and provide a detailed quantitative analysis of the lattice vibrations corroborated by reflectance measurements on single crystals. From these data, we determine the frequencies and damping coefficients of both LO and TO phonons and find strong broadening and anharmonicity of the lattice vibrations for all three halide perovskites. This indicates dynamic disorder of the lead-halide cage at room temperature, in agreement with recent low frequency, off-resonance Raman measurements of MAPbBr$_{3}$.\cite{yaffe_nature_2016} From our data we calculate the static dielectric constants of all three halide perovskites via the Cochran-Cowley relation.\cite{cochran_dielectric_1962} We use the derived dielectric constants and phonon frequencies to estimate electron-phonon coupling constants and polaron masses. To quantify the importance of optical vibrations for charge transport, we employ the Feynman polaron model at elevated temperatures in conjunction with exciton effective masses, which provides upper limits for the optical phonon-scattering limited charge carrier mobilities in methylammonium lead halide perovskites.

\section{Results}

\begin{figure}
\includegraphics[scale=1]{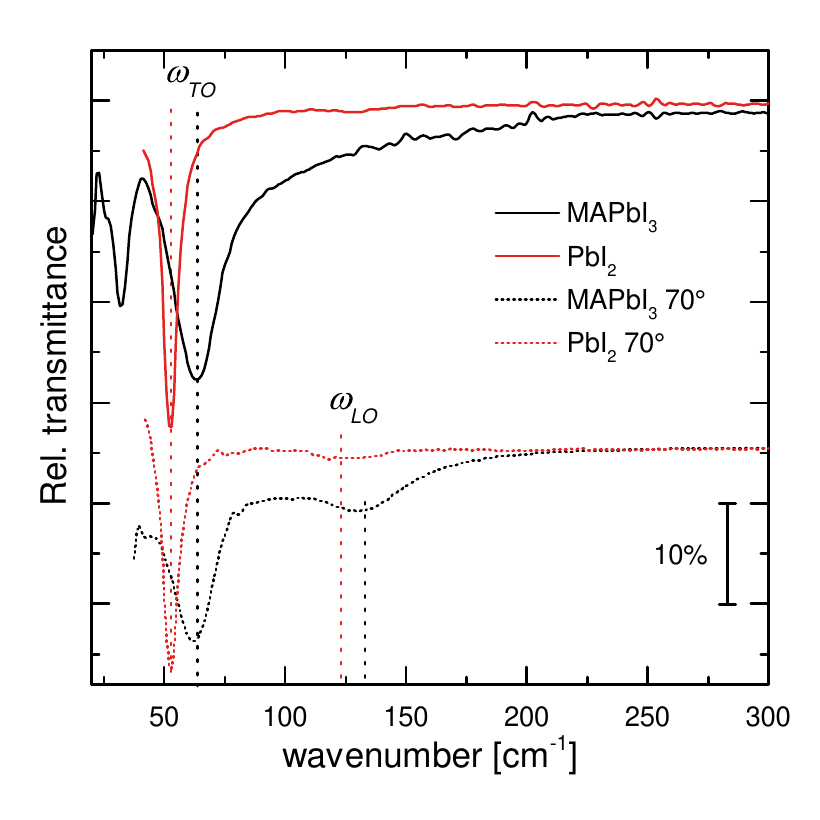}
\caption{\label{fig:1}
Relative transmittance of a 266\,nm thick MAPbI$_3$ (black) and a 130\,nm thick PbI$_2$ (red) thin film on a silicon substrate under normal incidence (solid), and under $70^{\circ}$ angle of incidence (dashed) with unpolarized light. The positions of the strongest respective TO and LO modes are marked with dashed vertical lines.}
\end{figure}

We measured far IR spectra in the range of $\unit[30-300]{cm^{-1}}$, which contain information on the Pb-X lattice vibrations, in contrast to the mid IR range that is dominated by molecular vibrations of the MA cation\cite{bakulin_real-time_2015,glaser_infrared_2015,muller_water_2015}. Figure \ref{fig:1} compares measurements of a MAPbI$_{3}$ thin film under two angles of incidence with the measurement of the corresponding lead halide, PbI$_{2}$. Whereas at normal incidence one can only determine the resonance frequency of the TO mode ($\omega_{\mathrm{TO}}$), it is possible to additionally determine the position of the longitudinal optical vibration ($\omega_{\mathrm{LO}}$) at non-normal incidence via the Berreman mode.\cite{berreman_infrared_1963} For polar materials such as MAPbI$_{3}$ the LO frequency is higher than the TO frequency, hence a clear LO-TO splitting occurs. The peak positions for MAPbI$_{3}$ and PbI$_{2}$, determined through detailed analysis of the data as discussed below, are indicated in Figure \ref{fig:1} and differ by only $\sim \unit[10]{cm^{-1}}$. This is reasonable given the similar bulk moduli $B$ for the two materials,\cite{rakita_mechanical_2015} which are directly related to the LO frequency of the crystals ($B\propto \omega_{\mathrm{LO}}^2$).\cite{aguado_prediction_2006} The strong broadening and asymmetry of the main TO mode of MAPbI$_{3}$ at 63\,cm$^{-1}$ (8 meV) is apparent when compared to the sharp and symmetric TO mode of PbI$_{2}$. Since MAPbI$_{3}$ is highly crystalline, as evidenced by typically very sharp XRD peaks at room temperature, the phonon broadening is unlikely to be explained by static disorder caused by the micro strain present in polycrystalline thin films, because the broadening is also present in single crystals as we show below. We rather attribute it to dynamic disorder of the perovskite structure,\cite{poglitsch_dynamic_1987,weller_complete_2015,frost_what_2016-1,yaffe_nature_2016,egger_hybrid_2016} similar to certain oxide perovskites.\cite{ostapchuk_polar_2001} Given the frequency range of our measurements and the similar peak positions of MAPbI$_{3}$ and PbI$_{2}$, this likely originates from disorder of the lead-halide framework that is coupled to fast reorientations of the methylammonium cation\cite{bakulin_real-time_2015,leguy_dynamics_2015}, and possibly accompanied by strongly anharmonic hopping of Pb and/or I.\cite{gervais_anharmonicity_1974,yaffe_nature_2016}

\begin{figure}[h!tb]
\includegraphics[scale=1]{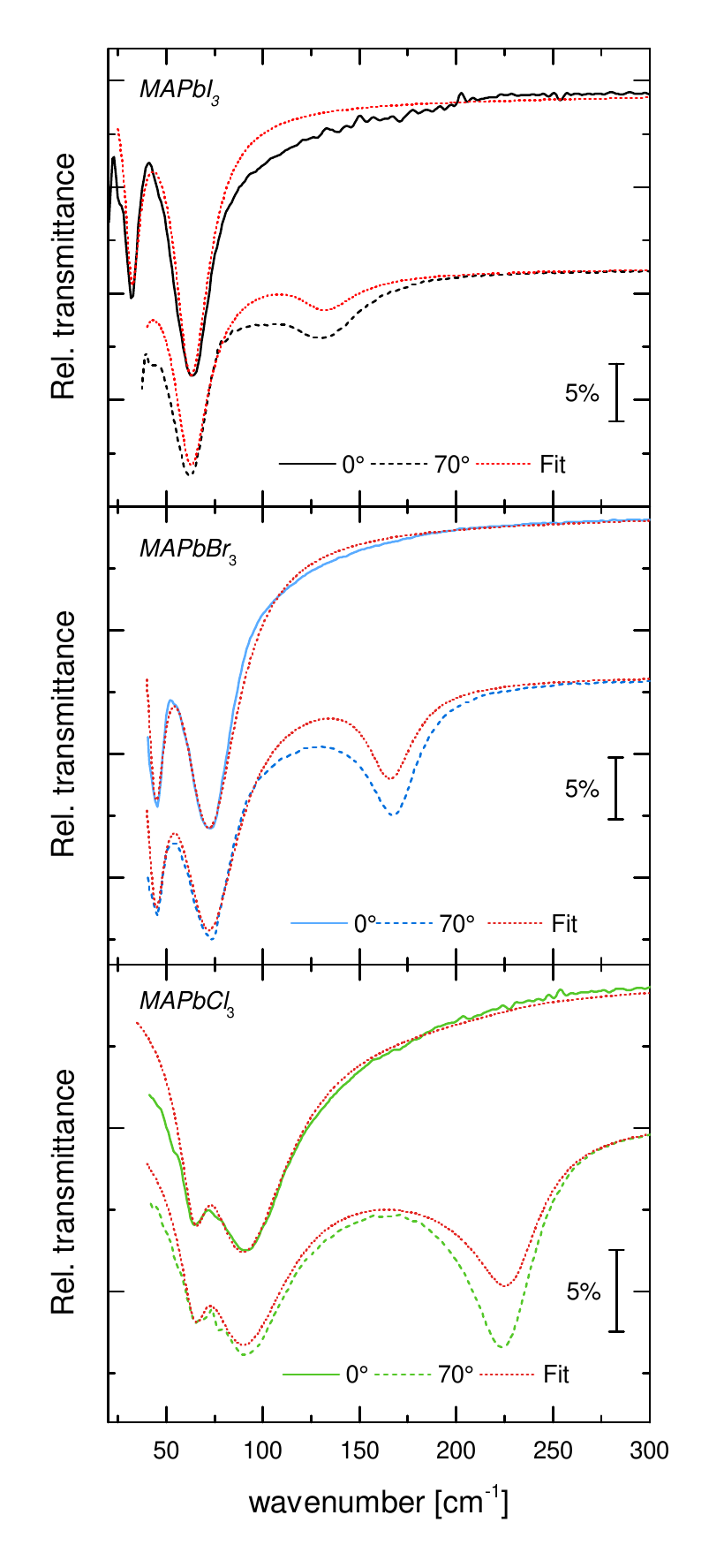}
\caption{\label{fig:2}
Relative transmittance of 266\,nm thick MAPbI$_{3}$ (black), 287\,nm thick MAPbBr$_{3}$ (blue) and 226\,nm thick MAPbCl$_{3}$ (green) thin films on silicon substrates at normal incidence (solid) and at $70^\circ$  angle of incidence (dashed) with unpolarized light. Model fits for both angles of incidence for every MAPbX$_{3 }$, are shown in red dashed lines, with the resulting dielctric function shown in Figure \ref{fig:4}.}
\end{figure}

Figure \ref{fig:2} shows the measured far IR spectra of all three methylammonium lead halides at two angles of incidence, together with optically modeled spectra. Given the strong LO-TO splitting and anharmonic phonon coupling apparent in our data, a harmonic potential of the vibrations cannot be assumed anymore. Therefore, we use the Gervais model\cite{gervais_anharmonicity_1974} to fit the spectra for the dielectric function, which extends the harmonic oscillator model to the case of strong oscillations and, in contrast to a Lorentz oscillator, allows for different damping coefficients for LO and TO phonons ($\gamma_{\mathrm{LO}},\gamma_{\mathrm{TO}}$):\cite{gervais_anharmonicity_1974,berreman_adjusting_1968}

\begin{equation}\label{eq:1}
\varepsilon(\omega)=\varepsilon_{\infty}\prod_{n} \frac{\omega_{\mathrm{LO,}n}^2-\omega^2+\mathrm{i}\gamma_{\mathrm{LO,}n}\omega}{\omega_{\mathrm{TO,}n}^2-\omega^2+\mathrm{i}\gamma_{\mathrm{TO,}n}\omega}.
\end{equation}

The model takes into account that TO modes appear as the poles of $\varepsilon(\omega)$, and LO modes as the zeros of $\varepsilon(\omega)$.\cite{gervais_anharmonicity_1974} The high frequency dielectric background $\varepsilon_{\infty}$ is taken from mid-infrared spectral modeling.\cite{glaser_infrared_2015} Two oscillators were used in equation \ref{eq:1} to model the spectra for each thin film, simultaneously for the two angles of incidence - one rather strong oscillator and a second weaker one on the low energy side of the spectrum. Although the fits show minor deviations from the measured data, the agreement is very good overall. Following previously published theoretical calculations,\cite{perez-osorio_vibrational_2015} we assign the mode at 32\,cm$^{-1}$ in MAPbI$_{3}$ mostly to a Pb-I-Pb rocking vibration, and the stronger mode at $\unit[63]{cm^{-1}}$ mostly to a Pb-I stretching vibration. A similar assignment of modes can be assumed for MAPbBr$_{3}$ and MAPbCl$_{3}$, given their chemical similarity to MAPbI$_{3}$.

\begin{figure}
\includegraphics[scale=1]{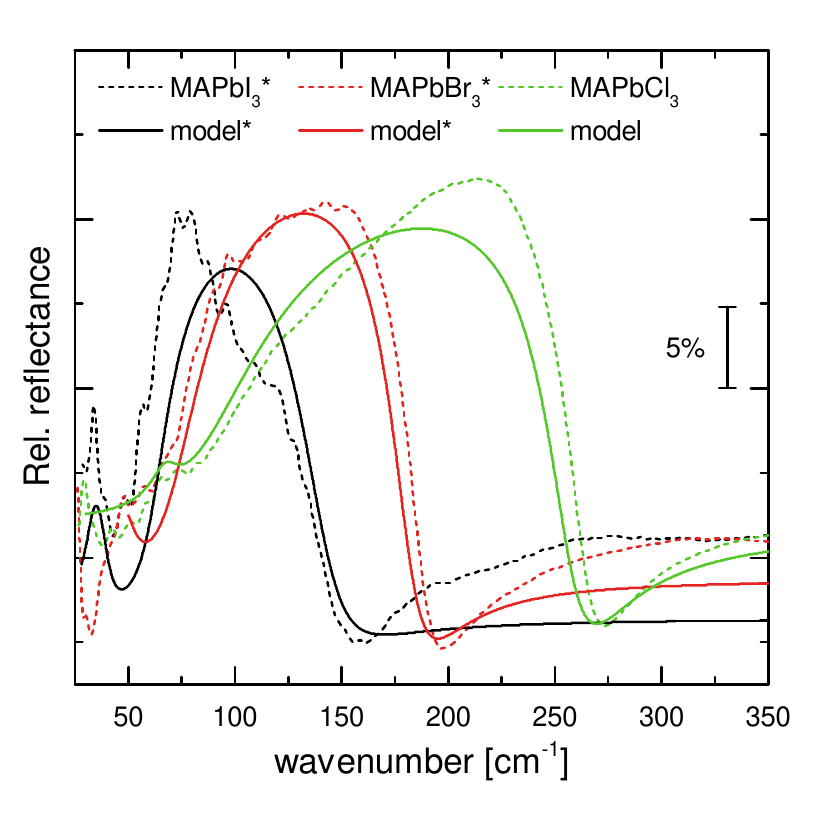}
\caption{\label{fig:3}
Relative reflectance spectra of MAPbI$_{3}$ (black), MAPbBr$_{3}$ (red), and MAPbCl$_{3}$ (green) single crystals at room temperature with unpolarized light under an $80^\circ$ angle of incidence (gold mirror as reference). The corresponding solid lines show calculated reflectance spectra using the dielectric function, derived for the thin films shown in Figure \ref{fig:4}. (Spectra with star were scaled in intensity).}
\end{figure}

We performed additional far IR measurements on single crystals to corroborate the thin film results and exclude possible influence of grain boundaries. We note that due to their thickness and extinction coefficients these have to be measured in reflection geometry and that the spectra of the I and Br single crystals were scaled in intensity because the crystal facets were smaller than the IR beam diameter. Figure \ref{fig:3} shows the measured reflection spectra of the single crystals together with simulated spectra that were calculated using the corresponding dielectric functions obtained from the thin film transmission measurements. All three halide perovskites display an increased reflectivity in the spectral range between the TO and LO modes. This so called Reststrahlen band is typical for polar materials with strong LO-TO splitting.\cite{rubens_versuche_1897,berreman_adjusting_1968} Due to the increasing $\omega_{\mathrm{TO}}$ and $\omega_{\mathrm{LO}}$
 values from I to Cl, also the Reststrahlen band is blueshifted, as can be seen in the single crystal measurements. Importantly, we observe very good agreement between the measurements and calculations for all three spectra, which underlines that the dielectric functions obtained from the thin film measurements are meaningful. 

\begin{figure*}
\includegraphics[scale=1]{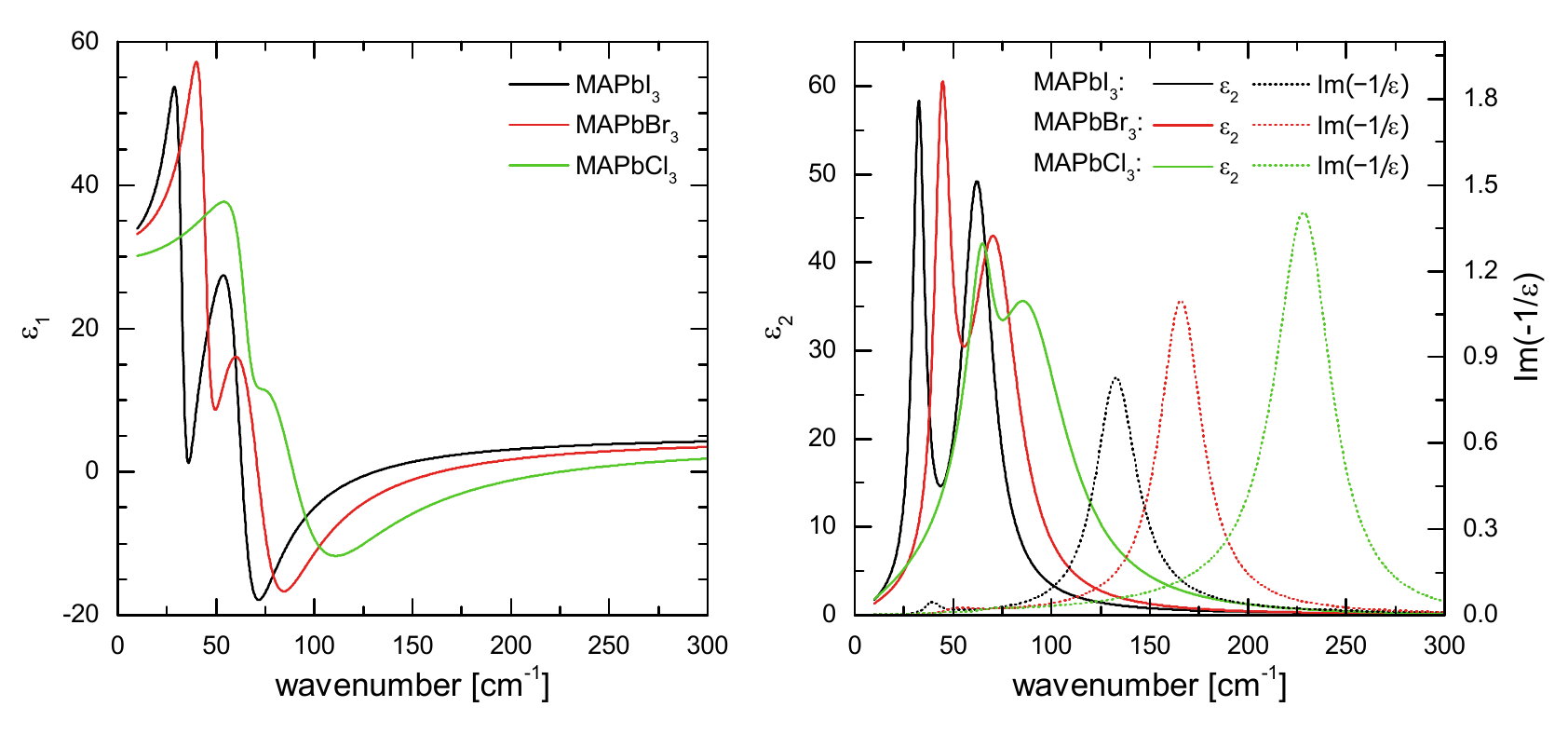}
\caption{\label{fig:4}
Derived dielectric functions for MAPbI$_{3}$ (black), MAPbBr$_{3}$ (red) and MAPbCl$_{3}$ (green). Left: real part $\varepsilon_1$ of the dielectric function. Right: imaginary part $\varepsilon_{2}$ together with the loss function $\mathrm{Im}(\nicefrac{-1}{\varepsilon(\omega)})$ (dashed lines). Peaks in $\varepsilon_{2}$ indicate positions of TO phonons, peaks in $\mathrm{Im}(\nicefrac{-1}{\varepsilon(\omega)})$ positions of LO phonons.}
\end{figure*}

The derived dielectric functions are shown in Figure \ref{fig:4}. The increase of the phonon frequencies in the I-Br-Cl series can be well described by the decrease of the effective ionic masses, $\mu_X$, of halides X involved in Pb-X oscillations $(\omega_{\mathrm{TO}} \propto \mu^{-0.5}; \mu_{\mathrm{I}}^{-0.5}:\mu_{\mathrm{Br}}^{-0.5}:\mu_{\mathrm{Cl}}^{-0.5}=1:1.16:1.61)$, which strongly corroborates our assignment of these modes to lead-halide stretch vibrations. Figure 4 also shows the loss function, $\mathrm{Im}(\nicefrac{-1}{\varepsilon(\omega)})$, the  peaks of which mark the positions of frequencies corresponding to LO phonons.\cite{berreman_infrared_1963}

\begin{table}
\caption{\label{tab:1}Summary of the obtained TO and LO phonon frequencies and dampings, as well as the high and low frequency dielectric constants.}
\begin{tabular}{ccccccc}
\hline
 & $\omega_{\mathrm{TO}}$  & $\gamma_{\mathrm{TO}}$ & $\omega_{\mathrm{LO}}$ & $\gamma_{\mathrm{LO}}$ & $\varepsilon_{\infty}$ & $\varepsilon_{\mathrm{Static}}$\\
  & [cm$^{-1}]$ & [cm$^{-1}$] & [cm$^{-1}$] & [cm$^{-1}$] &   &  \\

\hline

\multirow{2}{*}{MAPbI$_{3}$} & 32 & 9 & 40 & 11 & \multirow{2}{*}{5.0} & \multirow{2}{*}{33.5}\\
  & 63 & 20 & 133 & 30 & & \\

\multirow{2}{*}{MAPbBr$_{3}$} & 45 & 10 & 51 & 15 & \multirow{2}{*}{4.7} & \multirow{2}{*}{32.3}\\
  & 73 & 30 & 167 & 27 & & \\
\multirow{2}{*}{MAPbCl$_{3}$} & 66 & 15 & 70 & 16 & \multirow{2}{*}{4.0} & \multirow{2}{*}{29.8}\\
  & 89 & 52 & 225 & 34 & & \\

\hline
\end{tabular}
\end{table}

Table \ref{tab:1} summarizes the determined $\omega_{\mathrm{TO}}$ and $\omega_{\mathrm{LO}}$ frequencies for the different halide perovskites. Using the Cochran-Cowley relation,\cite{cochran_dielectric_1962}
\begin{equation}\label{eq:2}
\frac{\varepsilon_{\mathrm{Static}}}{\varepsilon_{\infty}}=\prod_{n} \frac{\omega_{\mathrm{LO,}n}^2}{\omega_{\mathrm{TO,}n}^2},
\end{equation}
we calculate the static dielectric constants for the different halide perovskites, also listed in Table \ref{tab:1}. Our derived values (33.5, 32.3, 29.8 for MAPb(I/Br/Cl)$_{3}$, respectively) are close to those published by Poglitsch and Weber\cite{poglitsch_dynamic_1987-1} (30.5, 39, and 26 respectively; measured at $f=90\,\mathrm{GHz}$). We note, however, that Onoda-Yamamuro et al.\ determined considerably higher values (45, 60, 58; measured at $f=1\,\mathrm{kHz}$).\cite{onoda-yamamuro_calorimetric_1990} The discrepancy could be related to mobile ionic species that can strongly influence impedance measurements.\cite{yang_significance_2015}

We now examine the implications of our results for electron-phonon interactions and charge transport. Fr\"{o}hlich described the movement of electrons in fields of polar lattice vibrations theoretically, introducing an interaction parameter\cite{biaggio_band_1997,frohlich_electrons_1954}
\begin{equation}\label{eq:3}
\alpha=\frac{1}{\varepsilon^*} \sqrt{\frac{R_y}{c h \omega_{\mathrm{LO}}}}\sqrt{\frac{m^*}{m_{\mathrm{e}}}}.
\end{equation}

The coupling constant $\alpha$ quantifies the interaction between a charge carrier with an effective mass $m^*$, and an optical phonon with frequency $\omega_{\mathrm{LO}}$ (in [cm$^{-1}$] units). The ionic screening parameter, $\nicefrac{1}{\varepsilon^*}=\nicefrac{1}{\varepsilon_{\infty}}-\nicefrac{1}{\varepsilon_{\mathrm{Static}}}$, is used to describe an effective dielectric background. $h$ is Planck's constant, $c$ is the speed of light, and $Ry$ the Rydberg energy. With multiple LO phonon branches, as is the case for the MAPbX$_{3}$ studied here (see Table \ref{tab:1}), an effective $\omega_{\mathrm{LO}}$ can be used, that is an average of the actual frequencies weighted by their spectral weight.\cite{hellwarth_mobility_1999} We note that this is also reasonable given the possibly strong anharmonicity of modes we find in our data, which may result in more complex scattering processes rather than interaction of carriers with discrete optical modes.

Within Fr\"{o}hlich's polaron theory, as extended by Feynman, the effective mass of the polaron $m_{\mathrm{p}}$ can be calculated as\cite{feynman_slow_1955}
\begin{equation}\label{eq:4}
m_{\mathrm{p}}=m^*\left(1+\frac{\alpha}{6}+\frac{\alpha^2}{40}+...\right).
\end{equation}
This means the polaron mass is higher than the effective mass obtained from band structure calculations due to electron-phonon interactions, thereby limiting the charge carrier mobility. We used $m^*=m_{\mathrm{ex}}^*$ with $m_{\mathrm{ex}}^*$ as the reduced exciton mass obtained from high quality magneto-optical measurements of MAPbI$_{3}$ and MAPbBr$_{3}$.\cite{galkowski_determination_2016} For MAPbCl$_{3}$, we performed density functional theory (DFT) based band-structure calculations using the VASP code,\cite{kresse_efficient_1996} the PBE exchange-correlation functional\cite{perdew_generalized_1996} and accounting for spin-orbit coupling (see Methods for details). We note that the reduced exciton effective mass obtained in this way is in good agreement with recent GW calculations,\cite{bokdam_role_2016} but individual electron and hole masses deviate more significantly.  Our calculated polaron masses, as well as the coupling constants $\alpha$, the ionic screening parameters $\nicefrac{1}{\varepsilon^*}$, and the corresponding polaron radii 
\begin{equation}\label{eq:5}
l_{\mathrm{p}}=\sqrt{\frac{h}{2c m_{\mathrm{ex}}^* \omega_{\mathrm{LO}}}}
\end{equation}
are listed in Table \ref{tab:2}. 

\begin{table*}
\caption{\label{tab:2}Determined polaron parameters and LO phonon  scattering limited charge carrier mobilities at room temperature for the three halide perovskites (this work) and GaAs for comparison.}
\begin{tabular}{ccccccc}
\hline
 & \parbox[t]{2cm}{effective\\mass$^a$} & \parbox[t]{2cm}{Ionic\\screening} & \parbox[t]{2cm}{Coupling\\constant} & \parbox[t]{2cm}{Polaron\\ mass} & \parbox[t]{2cm}{Polaron\\radius} & \parbox[t]{2cm}{Mobility\\$\mu$}\\
  & $\nicefrac{m_{\mathrm{ex}}^{*}}{m_{0}}$ & $\nicefrac{1}{\varepsilon^{*}}$ & $\alpha$ & $\nicefrac{m_\mathrm{p}}{m_{\mathrm{ex}}^{*}}$ & $l_\mathrm{p}$ [\AA] & $[\mathrm{cm}^2 \mathrm{V}^{-1}\mathrm{s}^{-1}]$ \\

\hline

MAPbI$_{3}$ & 0.104\cite{galkowski_determination_2016} & 0.17 & 1.72 & 1.36 & 51 & 197\\

MAPbBr$_{3}$ & 0.117\cite{galkowski_determination_2016} & 0.18 & 1.69 & 1.35 & 43 & 158\\

MAPbCl$_{3}$ & 0.20 & 0.22 & 2.17 & 1.48 & 27 & 58\\

GaAs\cite{adachi_gaas_1985} & 0.067 & 0.016 & 0.068 & 1.01 & 40 & 7000\\

\hline
\multicolumn{3}{c}{$^a$ in units of the free electron mass $m_0$}&&&&\\
\end{tabular}
\end{table*}

Importantly, these parameters can be used to estimate an upper limit for charge carrier mobilities $\mu$ in hybrid perovskites under the assumption that carriers are interacting only with optical phonons. Previously, Kadanoff's equation\cite{kadanoff_boltzmann_1963} for weakly coupled polarons in the low temperature regime was used to estimate the charge carrier mobility {\it } in MAPbI$_{3}$.\cite{la-o-vorakiat_phonon_2016} In view of the rather low Debye temperatures of MAPbX$_{3}$\cite{feng_mechanical_2014} and that for solar cell devices we are especially interested in $\mu$ at room temperature, we have to use the general result of Feynman et al.,\cite{feynman_slow_1955,feynman_mobility_1962} obtained in the absence of the low temperature approximation:\cite{biaggio_band_1997}

\begin{equation}\label{eq:6}
\mu=\frac{3\sqrt{\pi}e}{2\pi c\omega_{\mathrm{LO}}m_{\mathrm{ex}}^*\alpha}\frac{\sinh(\nicefrac{\beta}{2})}{\beta^{\nicefrac{5}{2}}}\frac{w^3}{v^3}\frac{1}{K}.
\end{equation}
Here, $\beta=\nicefrac{h c\omega_{\mathrm{LO}}}{k_{\mathrm{B}}T}$, {\it v}, and {\it w} are temperature-dependent variational parameters, and $K$ is a function of $\beta$, $v$ and $w$.\cite{biaggio_band_1997} We calculated {\it v} and {\it w} by minimizing the free polaron energy (see SI for details).\cite{biaggio_band_1997} The obtained upper limits for the charge carrier mobilties at room temperature are listed in Table \ref{tab:2}. We find values around 200, 150, and 50\,$\mathrm{cm}^2 \mathrm{V}^{-1}\mathrm{s}^{-1}$ for MAPbI$_{3}$, MAPbBr$_{3}$, and MAPbCl$_{3}$, respectivly. We estimate the error bar due to experimental uncertainties in the determination of the phonon frequencies as $\pm 30\,\mathrm{cm}^2 \mathrm{V}^{-1}\mathrm{s}^{-1}$.  The lower value for MAPbCl$_{3}$ is mainly a result of the higher $m_{\mathrm{ex}}^*$, which is related to the higher ionicity of the Pb-Cl bonds indicated by the higher LO-TO splitting.

\section{Discussion}
The observed strong broadening and anharmonicity of the lattice vibrations imply a large dynamic disorder in methylammonium lead halides at room temperature. The fast reorientation of the methylammonium cation in the PbI$_{3}$ host framework certainly contributes to this dynamic disorder.\cite{bakulin_real-time_2015,lee_role_2015,leguy_dynamics_2015,swainson_soft_2015-1} Moreover, it was recently reported that dynamic disorder is intrinsic to the general lead-halide perovskite structure even in the absence of organic cations.\cite{yaffe_nature_2016} This effect may well contribute to the broadening of the vibrations observed here. We note that the addition of a central peak to the dielectric function that would indicate strongly anharmonic hopping\cite{ostapchuk_polar_2001} of Pb or I, while consistent with our data, improves the quality of our fits only to a limited extent (see SI). Our obtained TO positions for MAPbI$_{3}$ are very similar to those previously published by La-o-vorakiat et al.\cite{la-o-vorakiat_phonon_2016} However, our oscillator strengths are roughly twice as large, possibly due to only partial surface coverage of the films used by La-o-vorakiat et al. (see SI for detailed comparison). We also find good qualitative agreement with a recently published dielectric function of MAPbI$_{3 }$ obtained from DFT and MD calculations.\cite{bokdam_role_2016} Our LO frequencies for MAPbI$_{3}$ and MAPbBr$_{3}$ are somewhat higher but still close to those indirectly estimated from temperature dependent photoluminescence measurements by Wright et al.,\ \cite{wright_electron-phonon_2016} confirming their analysis.

The determined polaron coupling constants for MAPbX$_{3}$ of $\alpha \sim2$ are at the higher end of the range typical for weak coupling (large polarons). Nevertheless, the polaron radii are consistent with large polarons, as they are far above the lattice constants of the materials and, in fact, show that polarons are spread over several unit cells. Consequently, the polaron masses are less than 50\% higher than the reduced excitonic mass determined from magneto optical studies.\cite{galkowski_determination_2016} Similar values are found by phonon calculations of MAPbI$_{3}$ leading to a coupling constant of 2.4 in the large polaron regime.\cite{brivio_lattice_2015-1,frost_what_2016-1}

A comparison of our mobility estimations to literature reports is complicated by the wide spread of published values. For example, for MAPbI$_{3}$ measured mobilities around 50\,cm$^{2}$V$^{-1}$s$^{-1}$ for thin films\cite{stoumpos_semiconducting_2013,wehrenfennig_charge-carrier_2014} and 2.5 or 165\,cm$^{2}$V$^{-1}$s$^{-1}$ for single crystals\cite{dong_electron-hole_2015,shi_low_2015-1} have been reported. Overall, our estimates are at the higher end of reported values, which is what one would expect as our numbers are upper limits in the complete absence of defect scattering, acoustic phonon scattering, and the possible impact of local polaronic distortions.\cite{neukirch_polaron_2016-1} Our findings nevertheless imply that room temperature mobilities well below 100\,cm$^{2}$V$^{-1}$s$^{-1}$ obtained from MAPbI$_{3}$ single crystal measurements may be strongly influenced by surface or interface effects. However, room temperature mobility values which are estimated to be significantly above $\unit[200]{cm^{2}V^{-1}s^{-1}}$ require careful scrutiny.\cite{valverde-chavez_intrinsic_2015}

We included the polaron parameters of GaAs\cite{adachi_gaas_1985} in Table \ref{tab:2} for comparison. We can attribute the large difference in mobility between GaAs and the perovskites to two material properties: (\textit{i}) the stronger ionic screening (indicative of a higher ionicity), and (\textit{ii}) the lower LO phonon energy in the perovskites. The lead halide perovskites are a rare example for semiconductors with a Debye temperature of the LO phonon frequency, $\Omega_{\mathrm{D}}=\nicefrac{h c\omega_{\mathrm{LO}}}{k_{\mathrm{B}}}$, below room temperature, leading to a strong occupation of phonon states at 300\,K which limits the mobility.

Finally, we briefly discuss the temperature dependence of the carrier mobility. For MAPbI$_{3}$, a comparison is complicated because the cubic-to-tetragonal phase transition temperature is close to room temperature. For MAPbBr$_{3}$, the transition occurs around 240\,K and so it is generally an easier task. In the SI we demonstrate numerically that between 350\,K to 250\,K the mobility of MAPbBr$_{3}$, including only the scattering due to optical phonons within the simple model used here, changes as $T^{-0.62}$. Indeed, in contrast to this finding it has been discussed that in the cubic phase the overall temperature dependence of the measured Hall mobility is best described as $T^{-1.4}$,\cite{yi_intrinsic_2016} the fingerprint of scattering due to acoustic phonons. \cite{milot_temperature-dependent_2015,oga_improved_2014,savenije_thermally_2014,yi_intrinsic_2016,brenner_are_2015} However, several theoretical reports showed that mobilities limited by acoustic phonon scattering should be extremely high,\cite{He2014,Wang2015} which also conflicts with the experimental data. Therefore, we are currently facing the contradiction that, as shown here, the large polaron model provides the incorrect temperature dependence but the correct magnitude of the mobility, and models including scattering only due to acoustic phonons follow the correct temperature dependence but strongly overestimate the mobility itself. This may indicate that one or several assumptions underlying these textbook models of charge-carrier scattering are actually not valid for the case of lead-halide perovskites. Indeed, as we have emphasized throughout this paper, lead-halide perovskites show strongly anharmonic vibrations and dynamic disorder.\cite{poglitsch_dynamic_1987,weller_complete_2015,frost_what_2016-1,yaffe_nature_2016,egger_hybrid_2016} Therefore, the canonical picture of charge carriers being scattered by harmonic vibrations of an otherwise rigid ionic lattice may not be valid for lead-halide perovskites. This situation calls for charge-carrier scattering models that do not start from the harmonic approximation and can take the intriguing structural dynamics of lead-halide perovskites fully into account. It also clearly highlights the need for more experimental data to better understand the transport of carriers in lead-halide perovskites.

\section{Summary}
In summary, we measured and analyzed quantitatively the lattice vibrations of the three methylammonium lead halide perovskites at room temperature. The spectra point to strong anharmonicity and dynamic disorder in these materials at room temperature, similar to certain oxide perovskites.\cite{ostapchuk_polar_2001} Our detailed analysis of the obtained far IR spectra allows us to directly determine the LO-TO splitting, and deduce the strength of electron-phonon coupling in MAPbX$_{3}$. Our estimated upper limit for the room temperature mobility in MAPbI$_{3}$ of $\sim\unit[200]{cm^{2}V^{-1}s^{-1}}$ indicates that there is still considerable room for further improvement of thin film quality, where mobilities of up to 50\,cm$^{2}$V$^{-1}$s$^{-1}$ have been achieved.\cite{stoumpos_semiconducting_2013,wehrenfennig_charge-carrier_2014} However, we also show that despite only moderate polaron masses, the low LO phonon frequency and high ionicity of the lead halide perovskites fundamentally limit the charge carrier mobilities in these materials.

\section{Experimental Methods}
{\it Thin Film Preparation}. Silicon substrates were cut and cleaned by sonication in acetone and isopropyl alcohol for 10 minutes. Afterwards they were dried in a N$_{2}$ stream. For perovskite film preparation we chose a vapor assisted evaporation process. Therefore we used as received PbCl$_{2}$, PbBr$_{2}$ (Alfa Aeser) and PbI$_{2}$ (Sigma Aldrich) for evaporation. We used a high vacuum chamber with a pressure of $1.5\cdot 10^{-5}$ mbar and evaporated 176\,nm thick layers of every one of those three materials at a rate of 8 nm/min. After evaporation the as evaporated films were transferred into a nitrogen filled glovebox for the vapor assisted process. Accordingly we put the lead halide samples in a petri dish and surrounded it with the corresponding methylammonium halide. We used as received MACl (VWR), MABr, and MAI (DYESOL). For perovskite formation we heated the petri dish for 2 hours at 165$^\circ$C, 160$^\circ$C and 155$^\circ$C respectively. Thus we achieved perovskite films with a layer thickness of approximately 300\,nm.

{\it Single Crystal Preparation.} Single crystals of MAPbCl$_{3}$, MAPbBr$_{3}$, and MAPbI$_{3}$ were prepared and characterized using the method described in the literature.\cite{Nayak2016} In figure S5 we show the optical images of the crystals that were used in this study.

{\it IR spectroscopy.} The as-fabricated perovskite thin films were transported in nitrogen atmosphere to a Bruker Vertex 80v Fourier transform IR (FTIR) spectrometer and measured under vacuum conditions ($\sim \unit[2]{mbar}$). Spectra were acquired using a liquid Helium cooled Si-Bolometer with unpolarized light and a resolution of 4\,cm$^{-1}$. All spectra were referenced to the bare Silicon substrate with natural oxide at the respective angle of incidence. Spectra of the single crystals were measured in reflectance under an angle of incidence of 80$^{\circ}$  using unpolarized light and a gold mirror as reference. Scaling of the single crystal spectra was executed due to an improper matching of the focal plane of reference and single crystal.
 
{\it Optical }{\it modeling}{\it :} Modeling of the measured thin-film spectra was performed using the commercially available software package SCOUT.\cite{theiss_scout_2015} The used optical model was made of a 1 mm thick Si substrate (described elsewhere\cite{glaser_infrared_2015}) and the MAPbX$_{3}$ film on top. The film thicknesses of the perovskite layer was determined by mid-infrared FTIR measurement of the same samples and fitting their thin film spectrum with the published optical model.\cite{glaser_infrared_2015} This thickness and the dielectric background $\varepsilon_{\infty}$ was used for the FIR optical model. The dielectric function of the perovskite thin films furthermore consisted of Gervais oscillators. The resonance frequencies of the transverse and longitudinal optical mode ($\omega_{\mathrm{TO}}$, $\omega_{\mathrm{LO}}$) and their damping ($\gamma_{\mathrm{TO}}, \gamma_{\mathrm{LO}}$) were fitted simultaneously for $0^\circ$  and $70^\circ$  angle of incidence.

{\it Computational methods:} DFT calculations were performed using a cubic unit-cell of MAPbCl$_3$ and a lattice constant of \unit[5.71]{{\AA}} as previously calculated, see Ref. \cite{Egger2014} for details. The plane-wave kinetic energy cutoff was set to 400\,eV and an $8\times 8\times 8$ $\Gamma$-centered $k$-point grid was used for self-consistently calculating the charge-density. The band-structure calculation was performed non-selfconsistently using an equally–spaced  $k$-grid of 100 points ($\Delta k \sim 0.01\, \mathrm{\AA}^{-1}$); the numerical convergence with respect to $\Delta k$ was verified. To estimate the reduced exciton effective mass, we have fitted the equation, $E(k)=E_0 \pm \frac{\hbar^2 k^2}{m_0 m_{\mathrm{e,h}}}$, to the onsets of the valence- and conduction band around $R$ in the direction of $\Gamma$ in the Brillouin zone, as the fundamental gap occurs at the $R$-point. The obtained reduced exciton mass is in good agreement with recent GW calculations for MAPbCl$_3$,\cite{bokdam_role_2016} although individual electron and hole masses show somewhat stronger deviations. This may imply that our estimate for the reduced exaction mass benefits from error cancellation. To test the reliability of our PBE+SOC results, we have also performed HSE+SOC calculations\cite{heyd_hybrid_2003,heyd_erratum:_2006} on carefully reduced $k$-grids, from which we obtained a virtually identical reduced exciton mass, similar to what has been reported for MAPbI$_3$ previously. \cite{menendez-proupin_self-consistent_2014}

\section{Acknowledgments}
We acknowledge the German Federal Ministry of Education and Research for financial support within the InterPhase project (FKZ 13N13656, 13N13657). The work in Oxford was funded by EPSRC, UK. M.S.\ and C.M.\ acknowledge financial support by the Heidelberg Graduate School of Fundamental Physics. P.K.N.\ is supported by Marie-Curie actions individual fellowships (grant agreement number 653184). D.A.E.\ and L.K.\ were supported by the Austrian Science Fund (FWF): J3608-N20 and by a research grant from Dana and Yossie Hollander, in the framework of the WIS Alternative sustainable Energy Initiative.

\providecommand*{\mcitethebibliography}{\thebibliography}
\csname @ifundefined\endcsname{endmcitethebibliography}
{\let\endmcitethebibliography\endthebibliography}{}

\bibliographystyle{rsc} 

\end{document}